# Resurgence of Electron Quantum Tunneling Sensors

Aishwaryadev Banerjee[1] (devaash88@gmail.com) and Carlos H. Mastrangelo (cmastran2009@gmail.com)[1]

[1]Department of Electrical and Computer Engineering, University of Utah, Salt Lake City, USA

**Abstract**

Quantum tunneling sensors are typically ultra-sensitive devices which have been specifically designed to convert a stimulus into an electronic signal using the wondrous principles of quantum mechanical tunneling. In the early 1990s, William Kaiser developed one of the first micromachined quantum tunneling sensors as part of his work with the Nasa Jet Propulsion Laboratory. Since then, there have been scattered attempts at utilizing this phenomenon for the development of a variety of physical and chemical sensors. Although these devices demonstrate unique characteristics such as high sensitivity, the principle of quantum tunneling often acts as a double-edged sword and is responsible for certain drawbacks of this sensor family. In this review, we briefly explain the underlying working principles of quantum tunneling and how they are used to design miniaturized quantum tunneling sensors. We then proceed to describe an overview of the various attempts at developing such sensors. Next, we discuss their current need and recent resurgence. Finally, we describe various advantages and shortcomings of these sensors and end this review with an insight into the potential of this technology and prospects.

1. **Introduction to Electron Quantum Tunneling**

Quantum tunneling is a phenomenon predominantly observed in nanoscale devices where a particle can pass or "tunnel" through a barrier potential greater than its energy. This is in direct contradiction with classical mechanics which states that any physical body cannot surmount a potential barrier greater than its own energy. Figure 1a. shows the schematic representation of an electron wave tunneling through a classically forbidden energy barrier.

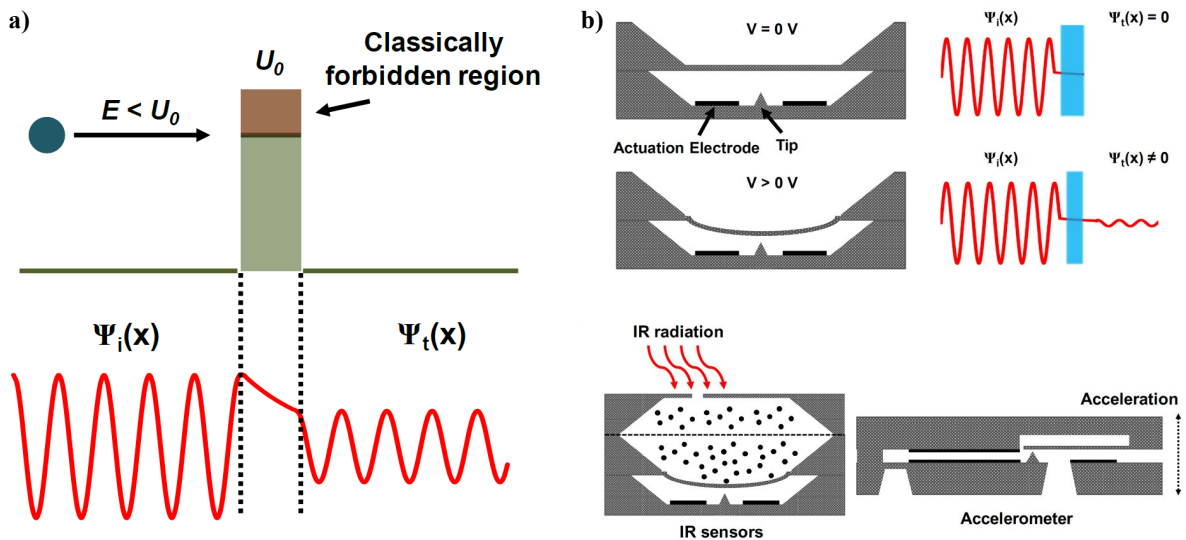

Fig 1: Schematic representation of quantum tunneling (a) of an electron wave across a energy barrier (b) early MEMS based quantum tunneling physical sensors.

The concept of quantum tunneling was a result of studies on radioactivity and one of its first occurrences was reported by Robert Francis Earhart, who in 1921 noticed an unexpected conduction regime when he was trying to understand conduction of gases between closely placed electrodes [1]. In 1926, Franz Rother, used a sensitive platform galvanometer to study field emission currents between closely spaced electrodes in high vacuum [2]. But it was in a paper published in 1927 by Friedrich Hund named "*Zur Deutung der Molekelspektren.*", roughly translated means "To the interpretation of the molecular spectra", where he discussed the phenomenon of quantum tunneling to explain the outer electron moving in atomic potential with two or more minima in potential energy separated by a classically impenetrable potential barrier [3]. The next major milestone in the history of quantum tunneling was achieved by Lothar Nordheim with the help of Ralph Fowler where he calculated the transmission probability of an electron wavefunction across a steep potential and showed that either reflection or transmission of the wave could occur with nonzero probabilities, whereas classically either one of the two can occur [4]. This is famously known as the Nordheim Fowler regime of electron tunneling where an electron wave tunnels through a triangular barrier under a high bias voltage. Later in 1927, Oppenheimer and also in March 1928, Fowler and Nordheim provided the analysis of transmission rate of an electron across a triangular barrier and proved the exponential dependence of tunneling probability on both barrier width and height. It was in the year 1930 that quantum tunneling proved most critical, when Oscar Rice interpreted the tunneling phenomenon as an analogy to alpha decay [5]. Rutherford, who was till then confused with the results he got from his scattering experiments (he observed that the α-particles which were emitted from U-238 were known to possess an energy of 4.2 MeV whereas the Coulomb potential was greater than 8.57 MeV), would eventually thank George Gamow, Ronald Gurney and Edward Condon who explained the phenomenon of alpha particle decay using quantum tunneling. About three decades later, Leo Esaki, Yuriko Kurose and Takashi Suzuki invented the tunnel diode (also known as the Esaki diode) which exploited the quantum tunneling phenomenon and displayed negative differential resistance (NDR) [6]. These were used as oscillators and high-frequency trigger devices and were noted for their extreme longevity. Devices fabricated in the 1950s are still functional. A few years later in 1963, John G. Simmons provided an analytical model describing tunneling current between two metal electrodes separated by a thin dielectric film [7]. This mathematical model has been widely used to describe tunneling current in solid-state devices and can also mathematically describe the working of quantum tunneling sensors. The modern-day world has much to thank Hund for his efforts in discovering the mystical phenomenon of quantum tunneling. Scanning Tunneling Microscopy (STM), Attenuated Total Reflection (ATR) spectroscopy method, Flash-memory are just some of the precious gifts of quantum tunneling.

2. **Electron Quantum tunneling sensors – generalized working principle, beginnings and a brief survey**

At the heart of the working principle of any electron quantum tunneling sensor (QTS) lies the exponential dependency of tunneling current density on potential energy barrier width and barrier height. Therefore, even a very slight change in either of these parameters results in an extremely high change in tunneling current density. All QTSs are therefore carefully designed so that upon detection of a specific stimulus, intended transduction leads to a change in barrier width/height which gives rise to a very high modulation of tunneling current. Accurate measurement of this current can therefore provide ultra-sensitive measurements of the causal stimulus. For example, in the IR-QTS designed by Kaiser and his team, tunneling current between a gold-coated silicon tip and a suspended membrane is measured as a function of membrane displacement, which in turn is caused due to expansion upon exposure to an oscillating IR source. Alternatively, one can measure the tunneling current, compare it to a reference value and apply correction signals to an electromechanical actuator in order to maintain a constant tunneling current value. This is achieved by applying forces to the moving sensor element so as to keep the tip–element gap constant (which is essentially the force-rebalance design). If the gain of the transducer, circuit, and actuator is sufficient, the tunneling current will be maintained in the presence of external disturbances. By monitoring the feedback signals produced by the control circuit, one can detect the signal forces applied to the sensor element. Figure 1b shows the working principle schematic of the some of the physical quantum tunneling sensors designed by Kaiser. The next section provides some details into the design and implementation of the first generation of QTSs.

Tunneling across a vacuum barrier was investigated by Binnig and Rohrer as part of their work which led to the invention of the Scanning Tunneling Microscope [8]. The device consisted of a piezo element which had a metal-tip fixed at the end. The tip was brought very close (in the z-direction) to the surface under observation which resulted in electrons tunneling across vacuum and a measurable tunneling current. A control-unit applied a voltage to the piezo element to maintain the tunneling current while the tip scans the surface in the x and y direction. Therefore, assuming the barrier height to remain constant, the applied voltage which was required to maintain the tunneling current was a function of the topography.

In the early 1990s, as part of his work with the Nasa Jet Propulsion Laboratory (JPL), Dr. William Kaiser developed a new class of micromachined sensor technology based on similar principles quantum tunneling. Taking inspiration from the STM, he wanted to exploit the exponential dependence of tunneling current on potential barrier height and tunneling distance. In principle, this would allow these sensors to display much higher sensitivities than their capacitive or piezoresistive counterparts. As part of this work, Dr. Kaiser built the world's first micromachined tunneling accelerometers and tunneling IR sensors [9]–[12] as shown in Figure 1b. Essentially, the accelerometer device consists of a tip-ended micromachined cantilever suspended on top of strategically placed electrodes. The cantilever is electrostatically pulled down by applying an appropriate bias voltage on one of the bottom electrodes. When the distance between the tip and the other lower electrode is near ~1 nm, tunneling current can be measured and detected by a feedback circuit. This feedback system controls the deflection voltage to maintain the position of the suspended cantilever. When the device experiences acceleration, the bias voltage (controlled by the feedback loop) maintains the relative position between the tip and the lower electrode. This bias voltage is recorded as output signal for the accelerometer.

The tunneling bolometer device consists of a small chamber filled with gas at atmospheric pressure. The chamber lies between a pair of thin $Si_3N_4$ membranes and consists of a suspended IR absorber membrane in the middle. Before exposure to IR radiation, a tunneling bias is applied to bring the deflection electrodes towards the tip. At a tip distance of ~1 nm, a small tunneling current of ~1.5 nA is measured. When IR radiation is incident on this device, it enters through the thin top membrane and gets absorbed by the membrane. This heats the air which expands and tries to lower the membrane to deflect closer to the tip, as shown in the schematic. A feedback system ensures that the tunneling current remains 1.5 nA by appropriately controlling the deflection voltage. This deflection voltage is the transducer output and is a measure of the IR radiation. Similar to such devices, Richard Colton built a tunneling magnetometer [13]. The operation of the device was similar to those described above. In this device, a magnetostrictive ribbon (Metglas 2605SC) was used to detect small changes in magnetic fields. The tunneling current between the ribbon and the metallic tip was maintained at a constant value by applying an appropriate displacement voltage. This voltage was the output signal for the magnetometer. The voltage required to maintain a constant tunneling current is essentially the "sensor output" and was used for measuring the appropriate stimulus.

These devices are some of the first and premier efforts in utilizing quantum tunneling for highly sensitive transduction and were mainly physical sensors which were used as motion-sensors or IR detectors. However, the physical structure of these devices made it extremely difficult to batch-fabricate them in the micro/nanoscale which restricted their use to highly specialized applications and they were therefore not deployed for general purpose. Also, these devices required an initial high bias voltage (greater than 120 V) for device operation. In addition to this, sensor operation also required a sophisticated feedback system to ensure proper working of the device. Finally, the working principle and device design also suggest that they might be susceptible to temperature fluctuations.

As an alternative to a suspended tip-based designs, one can also look towards nanogap electrodes for building tunneling sensors.

## 3. Nanogap electrodes for molecular electronics

Nanogap electrodes can be defined as a pair of electrodes, separated by a gap of just a few nanometers. These can be essentially considered as metal-insulator-metal (MIM) devices which have been widely used in the field of molecular electronics for mainly investigating the electrical properties of organic molecules. Although in principle, one can also realize a nanogap electrode system with semiconductors instead of metals. One of the first methods to realize a nanogap was the Mechanically Controllable Break junction (MCB) technique, which was used by M.A Reed when he was investigating the conductive properties of di-thiol molecules [14]. In an MCB junction, a notched metallic wire is glued to an elastic substrate. The substrate is bent with a piezoelectric element which causes the wire to fracture. After this, the distance between the two wire segments can be brought closer together by the piezo element to form a nanogap between the wires. Electrochemical methods are a simple way to fabricate nanogap electrodes. In this method, nanogaps are reversibly formed by controlled chemical deposition of specific atoms on lithographically defined nanogap electrodes to close down the gap between them. Dolan [15] established the oblique angle shadow evaporation method where an elevated mask in combination with an angled metal deposition is used to define metal leads with nanometer spacing. Electromigration, which has been infamously identified as a failure mode in the electronic industry, has been successfully used to fabricate nanogap electrodes [16]. In this method, a large current is passed through an e-beam lithographically (EBL) defined metal nanowire, which leads to electromigration of the metal atoms and eventual breakdown and fracture of the thin wire, leading to formation of a nanogap. Hatzor [17] introduced a new method where they used mercaptoalkanoic acids to pattern nanowires. Here, subsequent coatings of metal-organic resist on top of EBL metal patterns lead to a controlled gap between neighboring mercaptoalkanoic layers. Metal evaporation into the gap and the subsequent lift-off of the resist layer leads to a well-defined metal pattern, thereby forming a nanogap.

While nanogap electrodes fabricated by each of these methods have resulted in valuable results, which have gone a long way into understanding essentials of charge transport across break junctions, there are also certain fundamental disadvantages in using them [18]. For example, the MCB method is too cumbersome for high-density circuit applications since it requires macroscopic piezoelectric components for nanogap formation. Electrochemical methods require precise feedback mechanisms in real-time to monitor and accurately fabricate the electrodes with a precise nanogap between them. The oblique angle shadow method requires very low-temperature conditions for metal evaporation resulting in small metal grain sizes (thereby ensuring a uniform control of the nanogap between the electrodes). Electromigration essentially requires joule heating to form the nanogap, which means that there is also a high chance of undesired melting of metal. Also, electromigration sometimes leads to deposition of debris at the critical junction. Other methods involve expensive fabrication techniques like EBL and Molecular Beam Epitaxy (MBE) for consistent results. Most importantly, the nanogap features formed using these methods can be nonuniform in nature, which makes it difficult to use them for measuring tunneling current accurately.

To realize robust, uniform and CMOS compatible molecular tunnel junctions which can cater to detection of a variety of organic molecules, sidewall etched nanogap tunneling electrodes were introduced in 2006 [19]–[21]. Essentially, these devices consist of a top and bottom pair of electrodes electrically isolated by a thin insulating dielectric spacer layer, which is partially etched away along the edges. For using this structure as a chemical sensor, the device is then functionalized, and organic molecules end up covalently attached to the electrode pair at the nanogap site. These newly attached molecules provide additional electrical pathways for charge conduction between the electrodes. Therefore, this allows for inspection into charge transport across the molecular junction with and without conduction paths introduced by the foreign molecules, effectively decoupling the electrical characteristics of the covalently bonded molecules and the platform device. An example of such a nanogap system is illustrated in the schematic of Figure 2. The functional molecules (for example a SAM of thiols) can be localized to desired locations between the nanogap electrodes to form the metal-molecule-metal junctions.

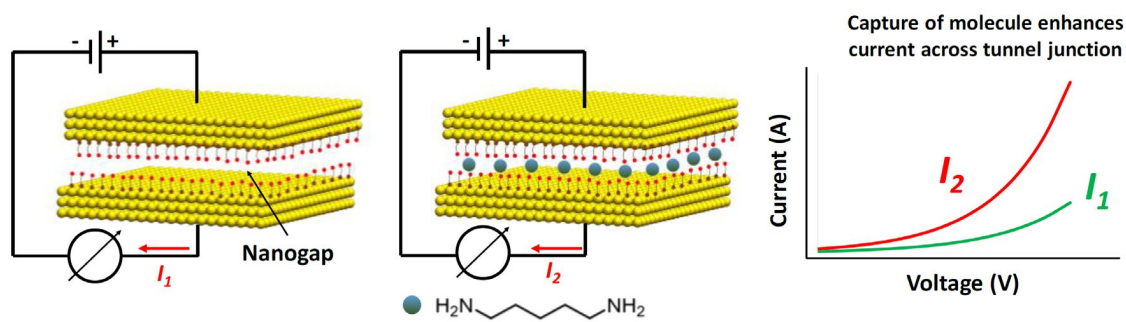

Fig 2: Schematic showing functionalized nanogap electrodes demonstrating augmented tunneling current after target capture.

Indeed, such nanogap electrode-based biosensors have been previously demonstrated in the field of molecular nanoelectronics. For example, Choi et al. fabricated an ultrasensitive nanogap junction for label free biomolecular detection and were able to detect DNA hybridization for 10 µM and 10 fM concentrations of target DNA [22], [23]. The authors also described a batch-fabrication process for parallel fabrication of sub-10-nm features [24], [25]. However, the fundamental working principle behind these nanogap sensors was detection of a change in dielectric constant upon successful detection of analyte within the nanogap metal-insulator-metal (MIM) structure. Using similar devices. Schlecht et al. developed a nanogap impedance biosensor for detection of thrombin using antibodies and RNA-aptamers as receptors [26], [27].

Previous efforts related to such nanogap electrodes were mainly confined for studying the electrical properties of organic molecules or for biosensing purposes. While these are extremely interesting devices, which can be batch-fabricated, and have provided noteworthy advancements in the field of molecular electronics, they do not completely exploit the fascinating properties of quantum tunneling (which was so wonderfully accomplished by Kaiser). Therefore, in order to realize highly-sensitive devices which can be batch-fabricated with reasonable uniformity and can be integrated with low-power sensor networks, Kim and Mastrangelo very recently demonstrated the use of nanogap electrodes as highly sensitive gas-sensors which consume very little power for continuous operation. Before describing these sensors in some detail, we will briefly describe the need for low-power sensors and sensor networks, and why current sensor technology cannot be utilized for this purpose.

4. Advancement of IoT and the need of ultra-low-power sensor networks

In today's world of ever-increasing power-hungry applications, inexpensive, portable devices which consume low power are crucial in building a truly interconnected and 'smart' civilization. With the advent of Internet of Things (IoT), we seem to be inching closer towards realizing such a society. The IoT is a concept which describes an interconnected system of physical devices which communicate with each other over the internet and can be remotely monitored and controlled [28]. The first internet connected appliance was a Coke vending machine at Carnegie Mellon University that could report if the drinks loaded inside the machine were cold or not. Since then, devices intended to be part of the IoT have evolved greatly. The number of devices which constitute the IoT network exceeded the world population way back in 2008. It is projected that this number will reach 50 billion in the year 2020 and the IoT market will surpass the market of the PC, tablet and phone combined. The potential growth in this industry is extremely high since only 0.06% of all possible devices have been optimized for IoT [29]. These devices include consumer electronics such as 'smart thermostats' which allow consumers to remotely control the temperature settings of their house as well as sophisticated MEMS sensors deployed in vehicles which provide driving assistance, optimized logistics and predictive maintenance. Since these devices must be remotely present across the IoT network grid, it is critical that they fulfill certain requirements. First and most importantly, they need to consume low power. Many IoT applications involve remote air quality monitoring and asset tracking. Such applications require the devices to be battery operated and deployed to remote areas of the world. Therefore, low power consumption is of utmost importance. Second, these devices also need to be highly portable. Consumer-electronic products such as the 'smart-watch' or applications such as remote-monitoring of pacemaker implants require the communicating devices to be as small as possible. Hence, large and bulky devices are not suitable for such applications. Third, these devices need to provide accurate and legitimate information for precise data collection and further analysis. For example, in remote sensing and medical applications, the employed devices cannot partake in faulty measurements. Since most of these devices are no longer implemented as part of 'stand-alone' functionalities, failure of one device will inevitably lead to a cascading failure of another. This domino effect needs to be prevented at all costs.

## 5. Current sensor technology and their drawbacks of w.r.t IoT applications

Existing sensor technology can be divided into mainly five categories: conductivity based, solid-state, optical, piezoelectric and polymer swelling induced gas sensors. Figure 3 shows the working principle of some of these sensors.

*Conductivity-based sensors*: In these types of sensors, the conductivity of the analyte-sensitive material changes when exposed to the analyte. Most common realizations of these devices include the use of conducting polymer composites [30], conducting polymers [31], [32] and metal oxides [31] as the sensing material. When the sensing layer comprised of a conductive polymer composite such as PEDOT:PSS or polypyrrole is exposed to the analyte, the polymer film absorbs the vapor which causes it to swell. This expansion causes a reduction in current conduction paths along the polymer which leads to an increase in resistance of the polymer film. Intrinsically conductive polymers such as polyaniline are also used as gas sensitive materials for such sensors. The principle of operation is the same as described above: exposure to a gaseous analyte leads to an expansion of the polymer which causes a change in electron density of the polymer chains and this changes the resistance of the polymer itself. Metal-oxide (MOX) based gas sensors work on the principle that the oxide (either p-type which respond to oxidizing gases or n-type which respond to reducing gases) reacts with the appropriate gas which leads to an excess of majority charge carriers (holes or electrons in p-type or n-type sensors). These excess charge carriers lead to augmented current conduction. Commonly used MOX sensors employ tin dioxide, zinc oxide, nickel oxide, cobalt oxide and iron (III) oxide as part of their sensing layer. These devices have certain inherent advantages such as conductive polymer-based sensors display reasonable selectivity in sensor response, they are relatively cheap to prepare and show linear response for a wide range of target analytes. MOX sensors show fast response and recovery times. However, the major disadvantage with these sensors is that they require high operating temperatures to function properly and hence cannot be used for low power applications.

*Solid-state gas sensors*: In these sensors, the threshold voltage of the semiconductor device [33], [34] (typically a MOSFET or PolFET) changes when exposed to a target analyte. This is because the interface of the catalytic metal (gate electrode) and the oxide layer gets polarized when exposed to the gaseous analyte and this changes the work-function of the metal and oxide layer. To facilitate a reaction between the analyte and the metal-insulator interface, usually a porous gas sensitive gate metal, such as Pd or a suspended gate design is used to provide access to the metal-insulator interface. These devices can be microfabricated using standard CMOS techniques which makes them highly compatible with existing CMOS circuitry and they are cheap to manufacture. However, the major drawbacks of these types of sensors are that they suffer from baseline drift and instability. They require complicated packaging and can function well only if the surrounding environment is controlled. Hence, they cannot be deployed for remote sensing purposes.

*Optical gas sensors*: In optical gas sensors, an optical fiber is coated with a fluorescent dye such as Pyranine or HPTS. This is encapsulated within a polymer matrix. When the optical fiber interacts with the target gas, the optical properties of the dye such as intensity, spectrum or wavelength change [35]. The sensitivity of the sensor depends on the type of dye and the polymer in which the dye is imbedded. Typically, adsorbents such as $Al_2O_3$ are often added to the polymer for improving detection limits [36]. The major advantages of these devices are that they have extremely fast response times and are immune to electromagnetic interference. However, to use them, one needs to implement complicated electronics and postprocessing software algorithms which makes them impractical for IoT based applications. Additionally, these sensors are also susceptible to photobleaching which can render them ineffective. Therefore, these sensors cannot be a part of the IoT framework.

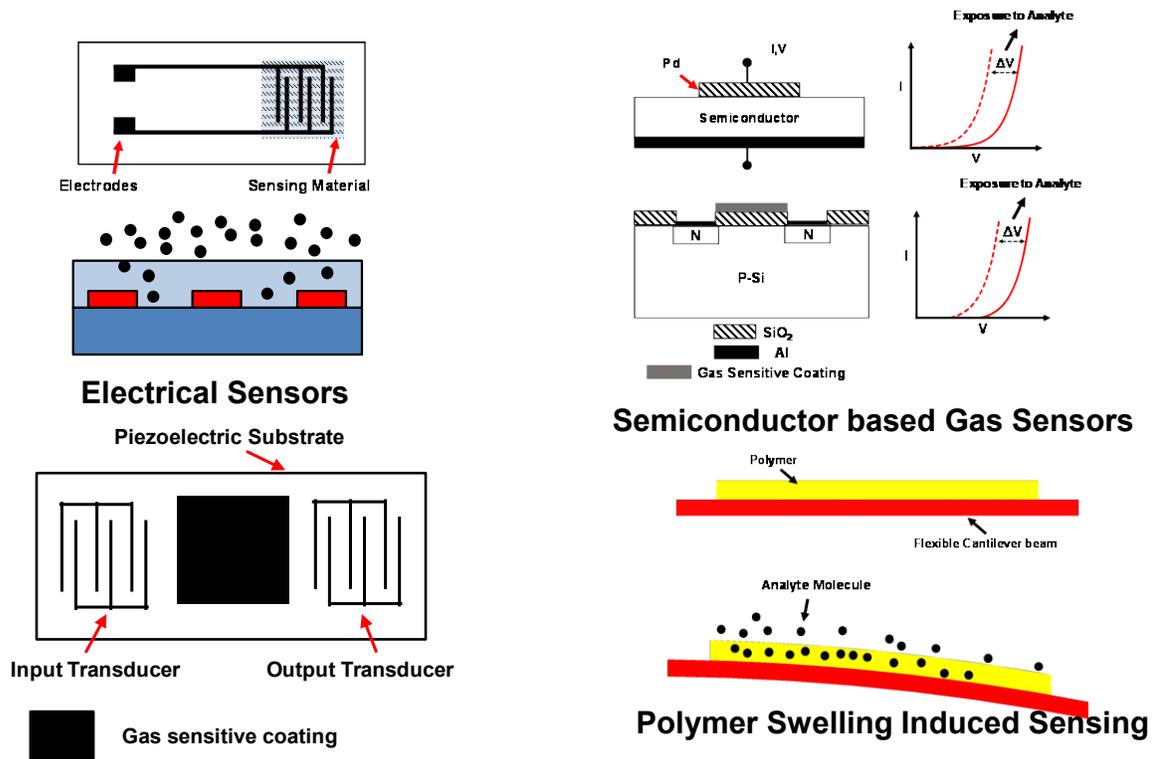

Fig 3: Different sensor types and their working principle.

*Piezoelectric gas sensors*: In piezoelectric gas sensors, the resonant frequency of the output signal changes when exposed to the target analyte. A typical surface acoustic wave (SAW) device includes two interdigitated transducers on a piezoelectric substrate (e.g. ZnO or Lithium niobate) with a polymeric gas sensitive coating in between [37]. An AC signal applied at the input electrode produces two-dimensional waves which travel along the surface of the substrate. When the device is exposed to the analyte, the coating adsorbs the analyte molecules and this change in mass of the gas sensitive coating changes the frequency of the traveling surface wave, which is then sensed at the output electrode. Similar to the SAW devices, quartz crystal microbalance (QCM) devices [38] operate on the principle that exposure to certain analytes leads to an increase in the mass of the gas sensitive material, which is deposited on a quartz crystal. This change in mass leads to a change in resonant frequency of the crystal and is used for detecting presence of analyte. The advantages of these devices are that they offer high sensitivity and fast response times. However, they require complex and expensive circuitry to function. Additionally, batch to batch reproducibility and low SNR add to the list of disadvantages of these devices.

***Polymer swelling induced gas sensors***: MEMS cantilevers have been widely used as gas sensors for more than a decade. In a typical MEMS-based humidity sensor [39], a Si free-standing cantilever is coated with a patch of polyimide, which is a moisture-sensitive polymer. The cantilever is suspended at a small distance on top of another electrode. When the device is exposed to change in ambient moisture, the polymer swells which causes a differential strain in the polymer-Si bimorph cantilever. This causes the cantilever to bend, thereby reducing the distance between the top cantilever and bottom electrode. The change in this distance causes the capacitance to increase which is a measure of the absorbed humidity. Such sensors have been widely used since they typically consume very low power and can be batch-fabricated. However, they are extremely temperature sensitive and display a lack of selectivity amongst different analytes [40]. Pattern recognition algorithms have been used to significantly improve the selectivity of these sensors. More examples of such sensors can be found here [41]–[46].

6. **Previous efforts in developing sensor technology for IoT applications**

Considering the various disadvantages mentioned in the preceding sections, most of the modern-day sensor technologies are not feasible to be used for IoT based applications. The commercially available gas sensors used for such purposes are used mostly for air-quality monitoring in manufacturing, agriculture and health industries and include devices which can detect analytes such as $CO_2$, $CO$, $H_2$, $O_3$ and $O_2$ [47]. These are usually electrochemical, photoionization and semiconductor-based sensors [48]. However, many of them require complicated circuits with multiple operational amplifiers for optimum performance. Correct sensor output is limited to operational temperatures $< 40°$ C and the devices are cross-sensitive to a multiple of commonly found gases.

There is considerable research going on in the field of low power gas sensing. Laubhan [49] proposed a low power IoT framework where they used multiple sensor chips as part of a wireless sensor network with configurable nodes which can be used to analyze motion detection, perform air quality inspection, measure humidity and temperature. Gogoi et al. [50], [51] also developed batch-fabricated multisensor platforms on a single chip which are probable candidates to be used for IoT based applications. Novel sensing mechanisms implemented by Chikkadi [52], Choi [53], Park [54], Woo [55] and Shim [56], have gone a long way at developing sensors for low-power applications. However, to more effectively develop sensor node systems for such applications, active research is being carried out to develop gas sensors which consume even lesser power and demonstrate higher selectivity/sensitivity. To accomplish this, very recently, a new family of nanogap gas-sensors which work on the fundamental principles of quantum tunneling has been developed. The following sections will briefly describe these devices and provide a commentary on their advantages and roadblocks to their realization as the next generation sensor for the IoT.

7. **Nanogap electrode-based Quantum Tunneling Gas Sensors**

As mentioned before, Mastrangelo and Kim present a new family of chemiresistive nanogap gas sensors that are based on electron quantum mechanical tunneling between two metal electrodes separated by nanogap junctions. As part of this research, they first developed a new batch-fabrication method of nanogap tunneling junctions and performed exhaustive characterization of the uniformity of the spacer layer, inspected tunneling current characteristics and determined the potential barrier of the thin spacer layer as well as maximum operating voltage for the device. After this, the nanogap electrodes are chemically functionalized by coating them with a self-assemble-monolayer (SAM) of thiol molecules. When the functionalized devices are exposed to the target molecules, they get "captured" by the SAM. The captured molecules form a molecular bridge across the junction producing an augmented electrical transport between the electrodes, along this bridge. This essentially reduces the barrier potential height for electron tunneling and exponentially increases the junction current density. This is the fundamental working principle of the nanogap electrode-based gas sensors and can be utilized for electrical detection of bridging target molecules. The significant increase in tunneling current upon target capture results in highly sensitive detection and very low leakage current prior to capture ensures extremely low power consumption. These devices are therefore potential candidates for low-power, portable and highly sensitive sensor applications. In this device, shown in the schematic of Figure 4, two gold electrodes separated by a ~6 nm-thick gap are coated with a self-assembled monolayer (SAM) of conjugated thiols. The resulting SAM-coated gap is designed such that if no gas is captured the tunneling resistance is high, in the order of $10^9$ Ω corresponding to the device's "OFF" state. If, however the SAM captures target gas molecules, the resulting metal-SAM-molecule-SAM-metal junction forms a molecular bridge between the otherwise electrically isolated electrodes. Figure 4 shows a schematic of the sensor action focusing on the device structure and target capture corresponding to an electrical switch which is "OFF" and then turns "ON", respectively.

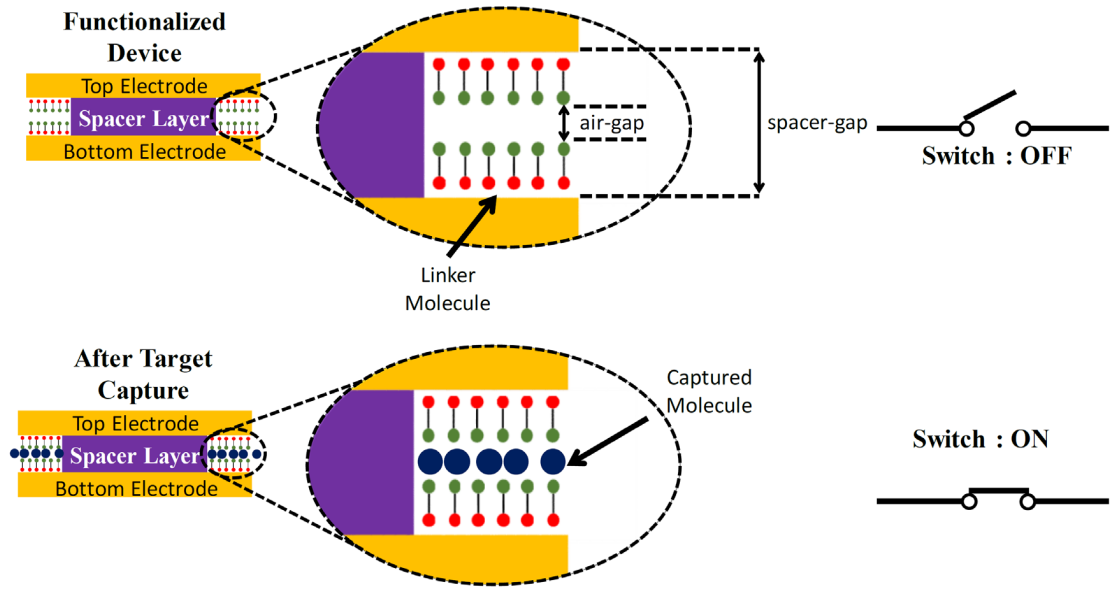

Fig 4: Schematic depicting the fundamental working principle of the nanogap electrode-based gas sensors.

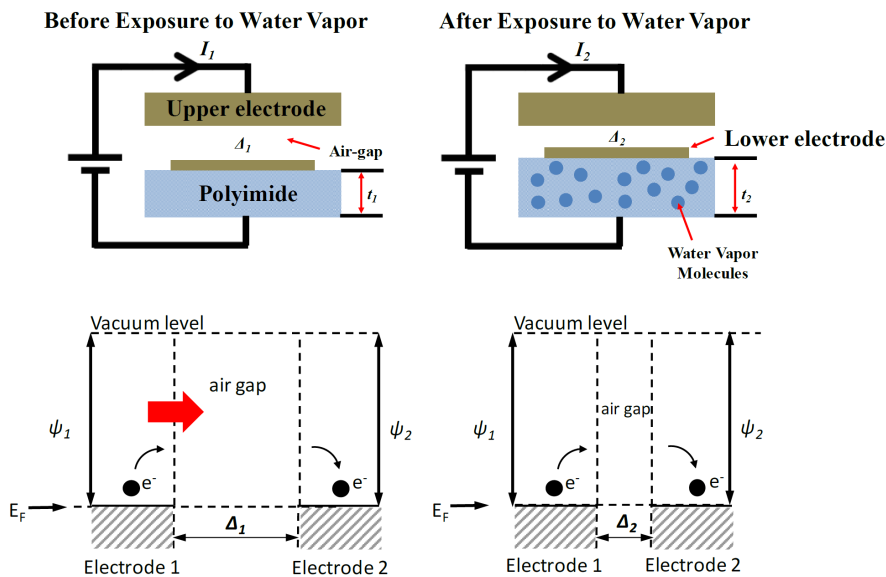

Fig 5: Fundamental working principle of quantum tunneling hygrometer.

In addition to these static nanogap electrode designs, they also presented the working of a new type of microfabricated quantum tunneling hygrometer that is able to provide large output range and a low temperature dependence. Figure 5 shows the schematic explaining the working principle of the humidity sensor. The device utilizes the expansion of a polymer that swells when exposed to humidity as in a polymer swelling based capacitive humidity sensor, but it produces a resistive output that measures the polymer expansion through tunneling current across a humidity dependent, thermally stabilized nanogap. The tunneling current changes many orders of magnitude providing similar output as the resistive type device but with a negligible temperature dependence. The following sections provide a brief description of this new family of gas sensors, with a focus on device design, fabrication, characterization and sensor performance.

### 7.1 *Batch fabrication of nanogap tunneling junctions: Device structure, design and fabrication*

The nanogap electrode assembly consists of a partially etched spacer film, sandwiched between two thin electrically isolated gold electrodes. The spacer is a sacrificial stack of a very thin dielectric layer of $SiO_2$ deposited using a plasma-enhanced ALD method, which provides excellent electrical insulation and an ultra-thin layer of sputtered α-Si, which acts as an adhesive layer between the top gold electrode and the dielectric material. A sacrificial plasma etch of the spacer layer creates a nanogap along the edges of the upper gold electrode. A schematic of the fabricated two-layer nanogap design is shown in Figure 6a. Since these nanogap devices will eventually be used for low-power and remote gas-sensing, it is essential that the leakage current during device operation be kept to a minimum so that the parasitic DC power-consumption is extremely low and since the junction leakage current is directly proportional to the overlap area of the electrodes, an electrode design with a low overlap area should generally ensure a lower leakage current. Therefore, the authors chose two design architectures having relatively low electrode overlap areas—a "square-overlap" layout (having an overlap area of ~16 μm$^2$), which is essentially a perpendicular arrangement of two thin metal wires and a "point-overlap" layout (having an overlap area of ~0.24 μm$^2$), which is a very low-overlap arrangement of lithographically patterned pointed tip-ends of patterned electrodes. This also allowed for measurement of current conduction characteristics as a function of the overlap area. Figure 6b-c shows the SEM images of the fabricated devices and nanogaps.

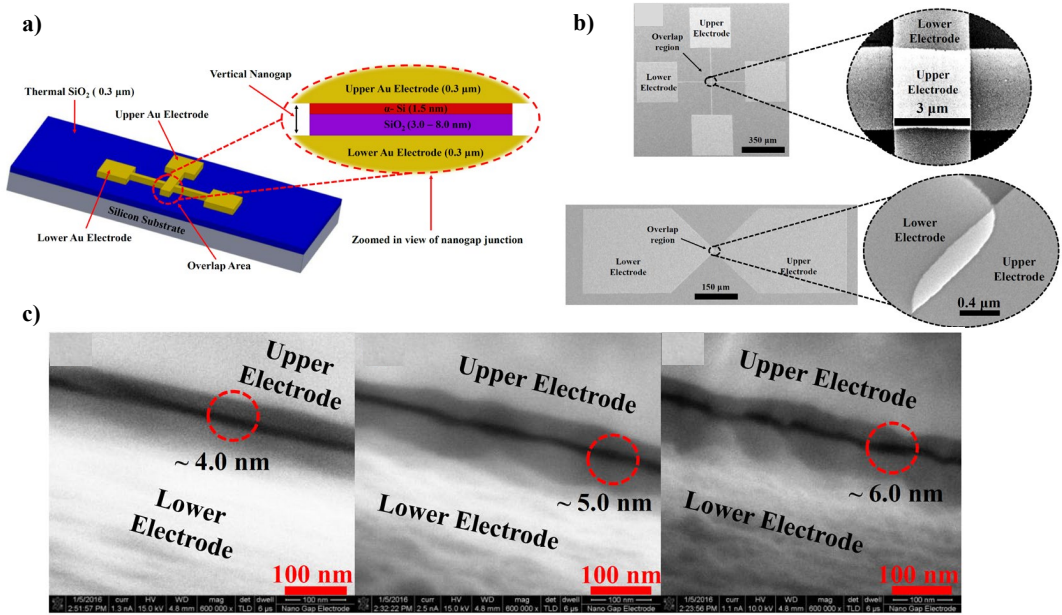

Fig 6 (a) Schematic of nanogap device structure and high-resolution SEM imaging of (b) top-view of the nanogap electrode pair and (c) nanometer-sized air-gap between the fabricated electrodes [57].

The fabrication process of the nanogap electrode assembly can be briefly explained as follows. They start by growing ~300 nm of $SiO_2$ on a Si wafer . This is followed by DC sputtering 25 nm of Cr and 200 nm of Au and subsequent patterning by traditional lithographic techniques to define the lower gold electrodes. The chemical solution Transene Au etchant TFA was used to selectively etch away the gold. Next, a desired thickness of dielectric material ($SiO_2$) was deposited for various time intervals, from 17 to 188 cycles of plasma-enhanced ALD process at a substrate temperature of 200 °C with the commercially available metal-organic precursors tris[dimethylamino]silane (3DMAS) on separate samples to fabricate nanogap electrodes with different spacer thicknesses. Then, an ultra-thin layer of α-Si was sputtered for 17 seconds at 50 W to get a ~1.5 nm film on each sample. Without breaking vacuum, another layer of ~200 nm of Au is sputtered and lithographically patterned to form the top electrodes. Finally, the samples are dry etched in an inductively coupled plasma etcher (Oxford 100 ICP) with $SF_6$ plasma for 40 seconds at an ICP forward power of 250 W with 45 sccm of $SF_6$ flow rate to partially remove the α-Si and $SiO_2$, thereby forming a nanogap along the edges of the top electrode.

After fabrication, a self-assembled-monolayer (SAM) linker solution is prepared by dissolving about 20 mg of the linker chemical in solution of about 4 ml dimethyl sulfoxide (DMSO) and 10 ml of ethanol. The fabricated devices were cleaned by treating them with $O_2$ RF plasma for one minute and then immediately immersed into the linker solution, to avoid any contamination. The exposed gold surface was functionalized by immersing the device in the proper linker solution for about 48 hours which ensured perfect formation of the SAM on all exposed gold surfaces, including those sandwiching the nanogap between them. After proper functionalization, the bare gold surface was coated by a self-assembled-monolayer of (4-((4-((4-mercaptophenyl)ethynyl)phenyl)ethynyl)benzoic acid).

### 7.2 *Sensor action*

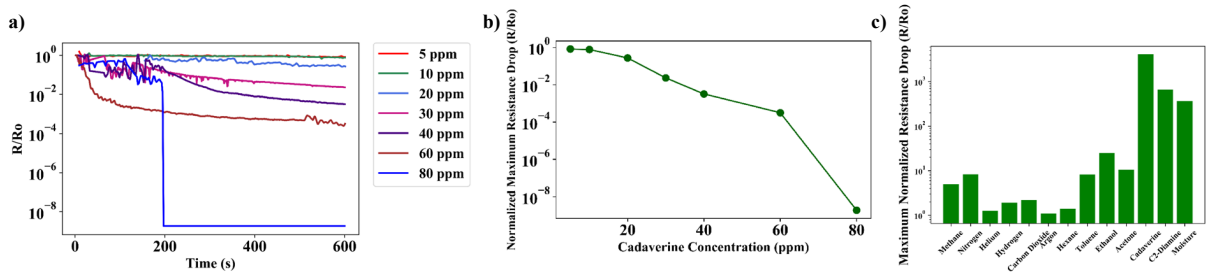

Fig 7: Sensor response against target analyte as a (a) function of time (b) various concentrations of analyte and (c) selectivity measurements [58].

Figure 7a shows the response of the nanogap device when exposed to different concentrations of cadaverine. As evident from the plot, exposure to 30-80 ppm of cadaverine lead to a reduction of junction resistance by ~ 2-8 orders of magnitude respectively. The standby power consumption of the sensor was measured to be less than 15.0 pW. If successful switching is defined to demonstrate an ON/OFF ratio of ~two orders of magnitude or more, we can conclude that the threshold for switching the nanogap devices 'ON' is ~30 ppm of analyte concentration in the testing chamber. Figure 7b shows the normalized maximum resistance drop of the nanogap sensor when it is exposed to different concentrations of cadaverine. To investigate the cross-sensitivity of the sensor, the device was also exposed to commonly encountered VOCs like acetone, ethanol and hexane as well as gases such as Helium, Nitrogen and $CO_2$. The authors define the sensor response as the normalized junction resistance drop after exposure to the analyte. Figure 7c shows the resistance ratio of the nanogap sensor when exposed to these analytes as compared to the ratio when exposed to the intended target gas, cadaverine. Measurements reveal a $R_{OFF}/R_{ON}$ ratio of more than four orders of magnitude when exposed to only 40 ppm concentration of cadaverine whereas a maximum $R_{OFF}/R_{ON}$ ratio of ~two orders of magnitude, when exposed to much higher concentrations of the other analytes. The concentration levels of the VOCs was maintained at greater than 10,000 ppm. To measure the device response in presence of other gases, we flooded the test chamber with the specific test gas and then monitored the resistance drop of the sensor. These results suggest a highly selective sensor action against most commonly found VOCs. More details regarding these devices can be found here [57]–[67]

### 7.3 Temperature compensated quantum tunneling hygrometer

In addition to sidewall-etched vertical nanogap quantum gas sensors, the authors also demonstrated a highly sensitive, temperature stabilized humidity sensor, working on the principles of quantum tunneling. The unique feature of this device was that in addition to ultra-high sensitivity and very low power consumption, the device also featured a passive temperature compensation scheme which ensured that the device would demonstrate an output signal only in the presence of changing humidity and not due to unwarranted temperature fluctuations.

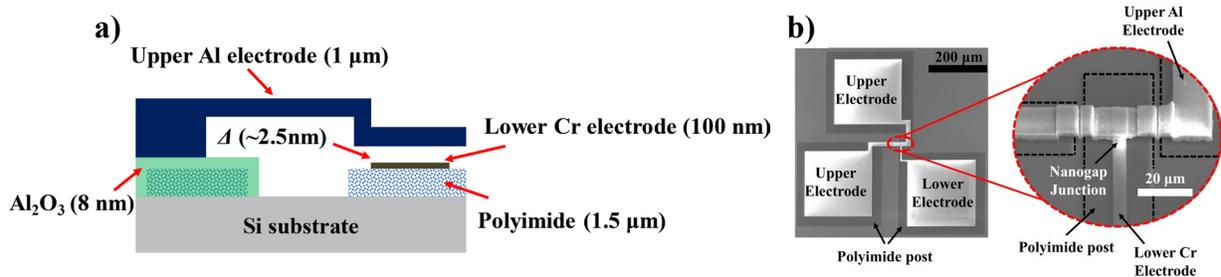

Fig 8: (a) Cross-section schematic of the tunneling hygrometer and (b) high-resolution SEM imaging of fabricated device [68].

*Working Principle*: The device consists of a pair of upper and lower electrodes, separated by an air-gap of ~2.5 nm as shown in Figure 5, illustrating the basic sensing mechanism.
The upper electrode rests at a fixed height. The bottom electrode rests on top of a hygroscopic polymer, polyimide that swells when humidity is absorbed; thus reducing the gap between the two electrodes. The swelling of the polyimide with humidity is linear corresponding to its humidity coefficient of expansion or CHE, which is 60-75 ppm/% RH [69]. Figure 5 also shows the electron band diagram across the nanogap. If the nanogap is very small electrons can tunnel from one electrode to the other establishing a conductive path. The magnitude of the tunneling current is exponentially dependent on the gap . In the simple configuration shown in the figure, the gap $\Delta$ is also affected by the thermal expansion of the polyimide, which makes the coefficient A and the resistance of the device strongly dependent of the ambient temperature.
To eliminate this strong temperature dependence, they utilize the differential device arrangement as shown in the schematic of Figure 8a where the two electrodes rest on polyimide patches of equal thickness, hence in the absence of humidity the nanogap distance is constant and independent of temperature fluctuations. However only one of these patches can absorb humidity; thus producing a humidity-induced nanogap change. The differential device assembly consists of an upper Al electrode and a lower Cr electrode separated by an air-gap of ~2.5 nm, standing on separate patches of 1.5 μm thick polyimide. As shown in Figure 8a, the polyimide patch under the upper electrode is covered with ~8 nm of ALD deposited $Al_2O_3$ diffusion barrier whereas that under the lower electrode is exposed to ambient atmosphere.
Thus, nanogap temperature stabilization was achieved by using cancelation of a common mode thermal expansion of both patches and humidity signal extraction by differential response to humidity between the two patches. Since both the top and bottom electrodes are standing on near identical polyimide patches, any increase in ambient temperature would lead to both the patches expanding almost equally. This ensures that in the event of temperature fluctuation, the nanogap distance between the top and bottom electrode will remain unchanged and the electrical response will be negligible in comparison to sensor response.

***Device fabrication***: The fabrication process started by growing ~1 μm of thermal $SiO_2$ on a 4-inch Si wafer. Polyimide was then diluted by dissolving uncured HD4104 polyimide (purchased from HD Microsystems) in N-Methyl-2-pyrrolidone (NMP) solvent in the ratio of 1:0.5 (wt/wt). This mixture was then dissolved by using a stirring it with a magnetic stirrer at 300 rpm for 2 hours to ensure perfect mixing of the polyimide and the NMP solvent. This mixture was then spin-coated on the sample at 2000 rpm following the standard procedures for polyimide processing. For curing the polyimide, the sample was kept in a nitrogenized environment oven for three hours at a temperature of 300 ° C. This procedure resulted in polyimide thickness of ~1.5 μm. Following this, we sputtered and patterned 200 nm of Al on the sample at 50 W and used it as a hard-mask to pattern the underlying polyimide. O2 plasma dry etching for 10 minutes at 100W was sufficient to remove the unwanted polyimide from the sample. Next, the remnant Al was stripped off by using commercially available aluminum etchant. Following this, we sputtered ~ 100 nm of Cr at 50 W on the sample and lithographically patterned it to define the lower electrodes. After this, we deposited about 2.5 μm of PECVD α-Si on the sample to form the sacrificial bridging supports for the upper electrode. This was followed by thermal ALD of ~8 nm $Al_2O_3$ and its lithographic patterning using BOE. Then, we sputtered ~2.5 nm of α-Si on the sample at 50 W to pattern the sacrificial spacer layer to define the thickness of the eventual nanogap between the electrodes. We next deposited ~1 μm of Al on the sample and patterned it to form the upper electrode. Finally, we sacrificially etched away the α-Si using $XeF_2$. A total of 1200 minutes of etching was required to completely etch away the sacrificial Si and release the upper electrode. SEM images of the fabricated device are shown in Figure 8b.

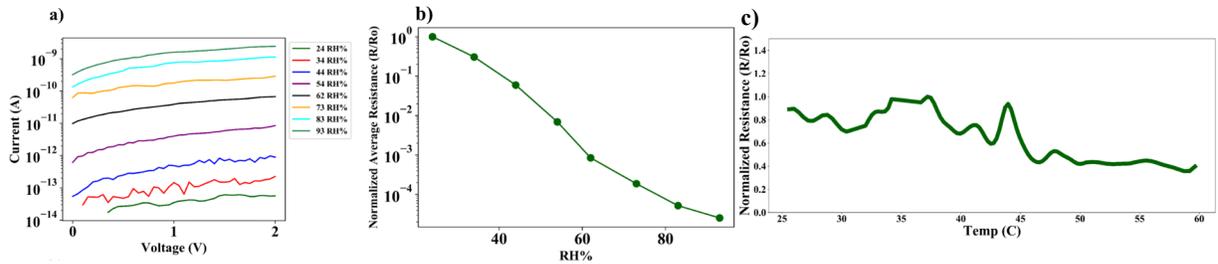

Fig 9 (a) – (b) Sensor for different RH% and (c) temperature response of the device.

***Sensor action***: After the device is fabricated, the upper and lower electrodes are separated by an air-gap of ~ 2.5 nm. Therefore, current flow across the nanogap junction involves conducting electrons tunneling through 2.5 nm of air-gap. However, when the device is exposed to an increase in humidity, only the un-protected patch of polyimide which is beneath the lower electrode absorbs water-vapor molecules and swells. This is because the $Al_2O_3$ acts as a diffusion barrier and prevents the polyimide patch under the upper electrode from absorbing water molecules [15]. This differential swelling of the polyimide patches results in the inter-electrode distance to reduce below 2.5 nm. Therefore, after absorption of water vapor molecules, the tunneling distance for the conduction electrons reduces. Since tunneling current exhibits exponential dependence on tunneling distance, increase in humidity levels lead to an exponential increase in tunneling current. In other words, junction resistance decreases exponentially when exposed to increased humidity levels.

Figure 9a shows the I-V curves of the device when exposed to an increasing RH% from ~20 – 90 RH%. As evident from the I-V plots, the increasing RH% leads to an increase in magnitude of current flow for the same voltage bias. Figure 9b is a plot of normalized average resistance vs RH%. These plots are a clear indication of exponential reduction of junction resistance with increasing RH%.

***Temperature compensation***: Figure 9c shows the temperature response of the device when subjected to a temperature change from 25-60 °C. As evident from the plot, the junction reduces <5 times when subjected to temperature changes. This is ~0.05% of the maximum resistance drop of the junction resistance when exposed to change in humidity. Therefore, it is clear that the device displays sufficient temperature compensation. Details of these tunneling hygrometers can be found here [68], [70]

## 8. Roadblocks

As mentioned above, the exponential dependence of tunneling current on potential barrier height/width which results in such high sensitivity of quantum tunneling sensors. However, it is this very dependency which prevents widespread usage of such sensors and restricts them to only specialized applications. Unavoidable variations due to fabrication errors can generally lead to non-uniformities of at least a couple of nanometers in thin-film dimensions, and by extension potential barrier dimensions as well. While this may seem negligible for most MEMS devices, which essentially operate in the micrometer scale, it can cause significant variation in quantum tunneling currents across the barrier which would inevitably lead to issues such as low repeatability and significant variation in device performance across a batch-fabricated set of devices.

For example, such issues have been often observed in nano-biosensors which comprise of either carbon nanotubes (CNT), nanoparticles, quantum dot structures, nanowires or nanorods. The unique properties of these nanomaterials/nanostructures can be exploited depending upon the specific analyte to be sensed [71]. For example, CNT's have been demonstrated to have an improved enzyme loading, ability to be functionalized, and low impedance for better electrical conduction. Quantum dots have been utilized for their excellent fluorescence properties and quantum confinement of charge carriers as well as tunable band energy. Nanoparticles aid in immobilization and enable better loading of bioanalyte as well as possess good analytic properties. Nanowires and nanorods are extensively used for biosensing purposes due to their compatibility with conventional MEMS processes as well as their high electrical conductivity. However, these nanosensors suffer from an alarmingly high varying response when exposed to identical analytes, which has been the most dominant roadblock in their path to practical and commercial usage. For example, when Choi et al. [72] developed a single-walled CNT (SWCNT) for detection of Staphylococcus aureus, the sensor demonstrated a significantly high average variance (standard deviation) in measured resistance change of the biosensor when reacting with Staphylococcus aureus. Figure 10a shows the resistance change of the SWCNT-based biosensors exposed to an increasing concentration of Staphylococcus aureus. As clearly evident from Figure 10a, the error bars in the sensor response plot covey a significant variance in sensor output for the same exposed concentration of Staphylococcus aureus and additionally, exposure to a varying concentration of Staphylococcus aureus also resulted in overlapping values of sensor output. Although most scientific articles offer limited explanation of this phenomenon, the root cause of the variance lies in the structural variations among CNTs synthesized in spite of being synthesized using identical methods. The three most common methods of preparing CNTs are arc discharge, laser ablation and CVD. Literature suggests that CNTs synthesized using these methods display an average diameter variance of ~30 nm. Additionally, there remains a significant structural variance within a single CNT. The working principle of most CNT based biosensors depends on the modulation of electron conduction through the CNTs upon interaction with target analytes. The electrical properties of CNTs are heavily dependent on the diameter and chirality of the specific CNT and they can be either metallic or semiconductors. Assuming a negligible voltage drop at the junction between two CNTs, the quantum tunneling conductance of CNT junctions formed between two carbon nanotubes can be approximated using the Landauer theory, where the conductance, G is given by:

$$G = \frac{e^2}{\pi \hbar} \sum_{m,n} |t_{mn}|^2 \quad (1)$$

where $t_{mn}$ is the transmission coefficient between the incident channel n and outgoing channel m. The exponential dependence of transmission probability in quantum tunneling assisted current conduction has been extensively discussed in the literature and the generalized expression for transmission coefficient can be written as:

$$t = \frac{1}{cosh^2(\beta L) + (\gamma/2)^2 \times sinh^2(\beta L)} \quad (2)$$

where L is the thickness of the barrier. In the context of current conduction through CNT junctions, the thickness of the barrier is approximately equivalent to the diameter of the CNT nanostructure and therefore, a variance in CNT diameter directly translates to an exponential variance in electrical characteristics of CNT junctions. Hence, a variance in CNT dimensions is most likely to cause a variance in CNT based biosensor output when exposed to same concentrations of target analyte. Indeed, there has been widespread reported variation in CNT based biosensors. Figure 10b-c shows the error bars in sensor output plots of various CNT based biosensors.

Similar variance in sensor outputs has also been observed in biosensors based on either nanoparticles, nanorods or quantum dots. For example, Ermini et al. developed peptide functionalized gold nanoparticles (AuNP) for detection of carcinoembryonic antigen in blood plasma [73]. The reported size of the synthesized nanoparticles was 30 nm ± 6 nm. The variance in the sensor output of NP based biosensors ranged from ~40% at lower analyte concentrations to ~20% at higher analyte concentrations. Since the current conduction mechanism between the aggregated nanoparticles is also quantum tunneling in nature, a variance in nanoparticle dimensions will also lead to an augmented variance in sensor output.

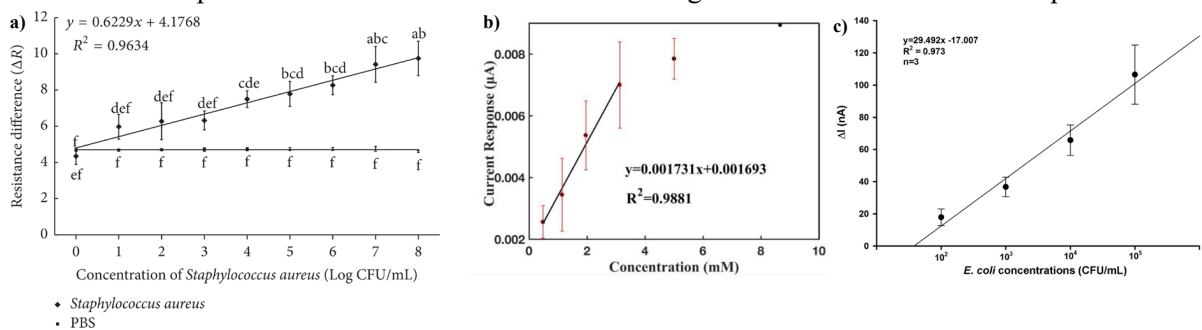

Fig 10: Sensor output of some CNT based biosensors showcasing the overlapping std. deviations for different analyte concentrations, thereby clearly demonstrating limited accuracy in such sensors [74]–[77].

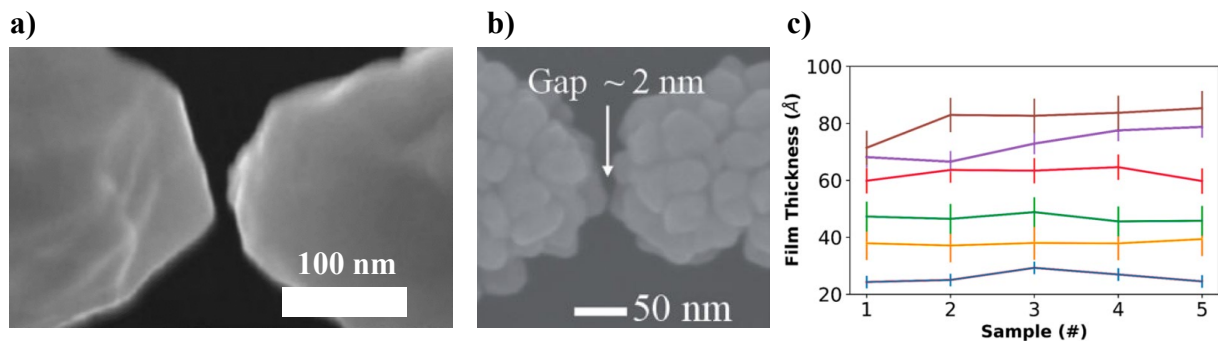

Fig 11: (a-b) High resolution SEM images of fabricated nanogap electrodes demonstrating the non-uniformity of nanogap dimensions and (c) variation in sacrificial layer thickness across multiple deposition cycles, also leading to nonuniformity in nanogap [78].

Additionally, biosensors based on batch-fabricated electrical nanogap devices have been developed in the last couple of decades for their excellent CMOS compatibility and their ability to enable direct electrical inspection of organic molecules. The gap between the electrodes can be suitably engineered to trap biomolecules of similar sizes which can be detected based on studying their electrical properties. To improve selectivity of the trapped biomolecules, these nanogap electrodes are often functionalized to enable specific binding of target analyte. Figure 11a-b shows the SEM of a fabricated nanogap electrode system. As evident from the SEM images, the dimensions of the fabricated nanogaps display significant non-uniformities. The dimensions of the gap between the electrode are crucial in determining its characteristics and performance. Due to the quantum tunneling nature of electron transport in these devices, non-uniformities in nanogap dimensions often lead to a significant variance in electrical characteristics of the fabricated nanogap electrodes. Banerjee et al. recently demonstrated the variance in electrical characteristics of batch-fabricated vertical nanogap tunneling junctions [57]. Figure 11c shows the variance in sacrificial layer deposition thicknesses which were used to define the dimensions of the nanogap between the electrodes. Therefore, it is but an inevitability that biosensors based on similar nanogap devices will also display some variation in sensor output. This variance in sensor performance and batch-to-batch irreproducibility presents a significant roadblock to the widespread and commercial use of quantum tunneling sensors.

## 9. Conclusion and future prospects

Quantum tunneling is one of the most intriguing phenomena observed in the natural world. Utilizing the exponential dependency of tunneling current on potential barrier height and barrier width, one can design sensors which can cause a change in either of the two parameters upon successful analyte detection, and in turn, cause an exponential change in tunneling current, which can be accordingly measured. The primary applications of such sensors include the development of very low power consuming and ultra-high sensitivity sensor and sensor systems. Traditionally, these devices had been developed for mainly specialized applications such as highly sensitive accelerometers/bolometers etc. Very recently, batch-fabricated nanogap electrode designs have been used to exploit these quantum properties and develop quantum tunneling gas sensors, which are extremely sensitive to analyte concentration change and consume very low power. The versatility of these devices have also been demonstrated when similar devices were functionalized with a different SAM and used to detect ultra-low concentrations of plant-emitted hormones [79], [80]. These devices are ideal for integration with a "wake-up" sensor network, where the active device only consumes electrical power in the presence of the analyte and remains in a dormant state otherwise.

However, as explained above, the quantum nature of current conduction in this device also poses significant challenges in its widespread usage. Unavoidable non-uniformities in current fabrication techniques inevitably lead to major variation in device performance and repeatability. Although, utilization of specialized techniques such as Pulsed Laser Deposition and Epitaxial Growth processes have been shown to deposit thin-films with remarkable uniformity, the process speed is highly reduced compared to other thin-film deposition techniques such as sputtering and evaporation. Therefore, for successful practical and widespread implementation of quantum gas sensors, one has to thoroughly optimize process parameters for device fabrication. Alternatively, one can also utilize artificial intelligence and machine learning based algorithms to develop a sensor system which compensate for the non-uniformities of the device properties. We feel that a combination of approaches including optimized process parameters and artificial intelligence will enable widespread utilization of quantum tunneling sensors and help usher in a new era of solid-state sensors with ultra-high sensitivity and very low power consumption features.